\documentclass[conference]{IEEEtran}
\IEEEoverridecommandlockouts
\usepackage{cite}
\usepackage{amsmath,amssymb,amsfonts}
\usepackage{algorithmic}
\usepackage{graphicx}
\usepackage{amsmath}
\usepackage{textcomp}
\usepackage{wrapfig}
\usepackage{xcolor}
\usepackage{hyperref}
\usepackage[super]{nth}
\usepackage{soul}
\usepackage[numbers]{natbib}

\usepackage {geometry}
\newgeometry {top=16.52mm, bottom=16.52mm, right=16.52mm, left=16.52mm} 

\def\BibTeX{{\rm B\kern-.05em{\sc i\kern-.025em b}\kern-.08em
    T\kern-.1667em\lower.7ex\hbox{E}\kern-.125emX}}
    
\usepackage{lipsum} % For some dummy text
\usepackage{filecontents} % To make the bib-file

\begin{document}
\bstctlcite{IEEEexample:BSTcontrol}

\title
  {
Maple: 
 A Processing Element for
 Row-Wise Product Based 
Sparse Tensor Accelerators
 \\ {\footnotesize \textsuperscript{*} A Preprint}
\thanks{* This paper was accepted at $60^{th}$ Design Automation Conference (DAC).}
}

\author{\IEEEauthorblockN{Midia Reshadi and David Gregg}
\IEEEauthorblockA{\textit{School of Computer Science and Statistics} \\
\textit{Lero, Trinity College Dublin}\\
Dublin 2, Ireland \\
Email: \{Midia.Reshadi, David.Gregg\}@tcd.ie}}
\maketitle

\begin{abstract}
Sparse tensor computing is a core computational part of numerous applications in areas such as data science, graph processing, and scientific computing. Sparse tensors offer the potential of skipping unnecessary computations caused by zero values. 
% But skipping zero values makes the computation and data access patterns much less regular than for dense tensors.
In this paper, we propose a new 
strategy for extending row-wise product sparse tensor accelerators. We propose a new
processing element called \textit{Maple} that uses multiple multiply-accumulate (MAC) units to exploit local clusters of non-zero values to increase parallelism and reduce data movement.
Maple works on the compressed sparse row (CSR) format and calculates only non-zero elements of the input matrices based on the sparsity pattern.
Furthermore, we may employ Maple as a basic building block in a variety of spatial tensor accelerators that operate based on a row-wise product approach.
As a proof of concept, we utilize Maple in two reference accelerators: Extensor and Matraptor. Our experiments show that using Maple in Matraptor and Extensor achieves 50\% and 60\% energy benefit and 15\% and 22\% speedup over the baseline designs, respectively. Employing Maple also results in 5.9$\times$ and 15.5$\times$ smaller area consumption in Matraptor and Extensor compared with the baseline structures, respectively.

\end{abstract}

% -----------------------Keywords------------------------------
\begin{IEEEkeywords}
sparse tensor accelerator, sparse computation, Gustavson's algorithm, row-wise product based technique, CSR compression
\end{IEEEkeywords}
% -----------------------Introduction-------------------------
\section{Introduction}
Sparse tensor algebra is used in applications such as  graph algorithms \cite{henry2007nodetrix}, scientific computing \cite{briggs2000multigrid}, and machine learning (ML) \cite{anandkumar2014tensor}. Sparse matrix-sparse
matrix multiplication (\textsc{sp}M\textsc{sp}M) often operates on very sparse matrices; for example, the web-Google (wg) matrix \cite{kolodziej2019suitesparse}\cite{suitesparse} consists of $916K\times916K$ elements with $5.1M$ non-zero elements and a density of $6.1e-6$. 
Highly-sparse computing tends to be memory-bound and leads to low energy efficiency on CPUs and GPUs \cite{duff2002overview}.
Hardware accelerators achieve energy efficiency by customizing the architecture to the problem. 

Three characteristics make sparse tensor accelerators efficient for sparse computing: (1) Employing compressed data formats such as CSR, CSC, or COO to reduce the amount of memory bandwidth required.
(2) Hardware support for vector intersection operations to match matrix inputs for multiply and accumulate.
(3) Parallelized computation using multiple processing elements.

Data reuse significantly alleviates the memory bottleneck in accelerators for dense data, such as deep neural network (DNN) accelerators. But data reuse is less easily exploited in accelerators for very sparse tensors, due to the large number of zero elements. For instance, inner-product dataflow \cite{hegde2018ucnn}\citep{qin2020sigma} maximizes output matrix reuse whereas, outer-product \cite{pal2018outerspace} maximizes the input matrix reuse and sacrifices output matrix reuse. Inner-product is inefficient with highly sparse matrices and outer-product suffers from merging large partial output matrices.
Gustavson’s algorithm or row-wise product multiplication (row times row) is arguably the most efficient dataflow strategy for \textsc{sp}M\textsc{sp}M accelerators; it is also widely used in CPUs and GPUs \cite{kjolstad2017tensor}\cite{wang2014intel}. 

The main objective of this work is to improve energy efficiency by reducing data movement between memory levels in Gustavson-based tensor accelerators.  Our strategy is to increase the amount of local computation inside each PE.
To this end, we propose a processing element called \textit{Maple} (\underline{M}ultiply and \underline{A}ccumulate in \underline{P}aral\underline{le}l), comprising several MAC units to enhance the local computation and minimize data movement between PE and higher level memory elements. 

Maple may be used as a key building block in spatial Gustavson-based accelerators.
The modular structure of Maple helps designers to use it with low redesign overhead. 
Maple operates directly on CSR-based compressed data and uses metadata to perform intersection and MAC operations only on non-zero elements. Therefore, there is no need to use separate logic in the input and output ports of the Maple PE to perform intersection and the CSR decompression functions.

To evaluate our solution, we utilize Maple in two state-of-the-art sparse tensor accelerators that work based on Gustavson's algorithm: Extensor \cite{hegde2019extensor} and Matraptor \cite{srivastava2020matraptor}.
Our evaluation process is based on 45nm technology using various sparse benchmarks. Our experiments demonstrate significant energy and area benefits over the baseline configurations. Additionally, employing Maple achieves 15\% and 22\% speedup in Matraptor and Extensor accelerators, respectively.

Overall, we make the following contributions:
\begin{itemize}
\item We propose a \textit{Maple} processing element consisting of several MAC units to parallelize computations within each PE and exploit the locality of reference in computing and accumulating partial sums in sparse matrix multiplication. 

\item \textit{Maple} operates based on the CSR data structure, which may be used as a PE building block in any spatial Gustavson-based tensor accelerators.

\item Employing \textit{Maple} in sparse tensor accelerators significantly reduces the energy consumption owing to local processing. \textit{Maple}'s customized design gains substantial area benefit compared to utilizing baseline PE. Meanwhile, the speedup is also achieved due to the high parallelism attained by employing multiple MAC units. 

\end{itemize}

% -----------------------Background-------------------------
\section{Background}
% -----------------------Tensor Terminology------------------
\subsection{Tensor Terminology}
Tensors are $N$-dimensional arrays that are represented by $N$-tensors notation. For example, scalars are 0-tensors, vectors are 1-tensors, and 2-tensors are matrices. 
We denote the matrices by upper case letters (e.g., \textbf{A} and \textbf{B}), where $\textbf{A}[i,j]$ represents the matrix $\textbf{A}_{m\times k}$ (2-tensor) with $m\times k$ size 
% and $\textbf{A}\in \mathbb{R}^{i\times j}$ 
in breve notation.
$\textbf{A}[a : b, x : y]$ represents a slice of the matrix \textbf{A} in the range of $a$ to $b$ in row and $x$ to $y$ in column and $\textbf{A}[i, : ]$ signifies an unbounded \textbf{A} slice in columns that contains all elements in row $i$ of the \textbf{A} matrix \cite{yesil2022dense}.

\begin{figure}[]
\centering
\centerline{\includegraphics[scale=0.32]{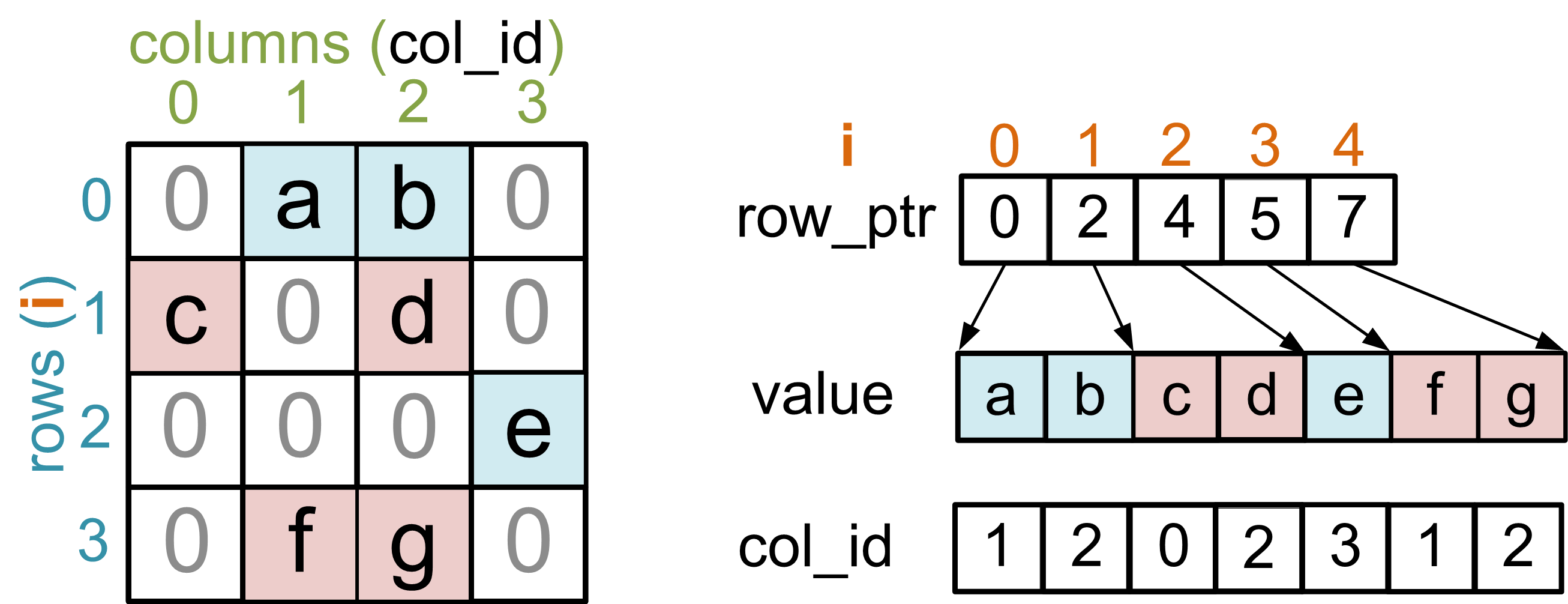}}
\caption{Compressed Sparse Row (CSR) data structure of matrix \textbf{A}.}
\label{fig.1}
\end{figure}
% -----------------------Sparse Data Structure-------------------------
\subsection{Sparse Data Structure}
Compressed data formats are used to avoid storing zero data values in memory for large sparse matrices. Several compression formats have been proposed in order to reduce storage space while maintaining high-performance non-zero data access. 
Compressed formats contain nonzero values called \textit{data} and additional information about the coordinates of nonzero values called \textit{metadata}.

The compressed sparse row format (CSR) is the most widely used compression format in sparse computing platforms. It represents a sparse matrix using three vectors: {\fontfamily{lmss}\selectfont value}, {\fontfamily{lmss}\selectfont col\_id}, and {\fontfamily{lmss}\selectfont row\_ptr}. The {\fontfamily{lmss}\selectfont value} vector comprises all nonzero elements and the {\fontfamily{lmss}\selectfont col\_id} vector stores the column coordinate of each element of the {\fontfamily{lmss}\selectfont value} vector.
For each row index {\fontfamily{lmss}\selectfont $i$}, {\fontfamily{lmss}\selectfont row\_ptr[$i$]} stores the starting location of the $i^{th}$ row in the value tensor. 

In this paper, we use simple symbols to represent the various parts of the CSR data structure. 
For example, according to Fig. \ref{fig.1}, {\fontfamily{lmss}\selectfont A.value[$i$]} for {\fontfamily{lmss}\selectfont $i$=0}   is equal to the non-zero values at the first row of the matrix \textbf{A} that are {\fontfamily{lmss}\selectfont \{a,b\}} and {\fontfamily{lmss}\selectfont A.col\_id[$i$]} is equal to {\fontfamily{lmss}\selectfont \{1,2\}}. We more specifically identify the non-zero element by $A.value[0][1]=a$ that means the non-zero element in the first row ($i=0$) and $col\_id=1$ of the matrix \textbf{A}.

% -----------------------Row-wise (Gustavson) product-based hardware accelerator-------------------------
\subsection{Row-wise product-based accelerator}
The row-wise product or Gustavson-based accelerator is spatial hardware dedicated to matrix multiplication that supports sparse matrices by skipping operations with zero values. 
The accelerator has a multi-level memory organization and takes advantage of the memory hierarchy. As a result, the datapath comprises several processing and storage elements arranged in two dimensions.

There are two levels of storage elements (SE) named $L_1$ and $L_0$; each level is arranged in two dimensions. $L_1$ SEs may be based on scratch pad memory (SPM) and $L_0$ SEs are registers inside each PE that are connected to a multiply and accumulate (MAC) unit.

For example, \textit{Matraptor} has two levels of memory including two storage elements in $L_1$ called \underline{sp}arse matrix \underline{A} \underline{l}oader (SpAL) and \underline{sp}arse matrix \underline{B} \underline{l}oader (SpBL), and sorting queue buffers inside PEs as $L_0$. All storage elements in each level are arranged in a $1 \times n$ array. 

\textit{Extensor} has two memory levels: \underline{l}ast-\underline{l}evel \underline{b}uffer (LLB) and \underline{p}artial \underline{o}utput \underline{b}uffer (POB), both of them are storage elements in $L_1$. Meanwhile, the internal \underline{PE} level \underline{b}uffers (PEB) are defined as $L_0$.
Another example is G\textsc{amma} \cite{zhang2021gamma}, which has two memory levels:  the Fiber buffer is $L_1$ and the PE's internal registers are $L_0$.
\begin{figure}[]
\centering
\centerline{\includegraphics[scale=0.43]{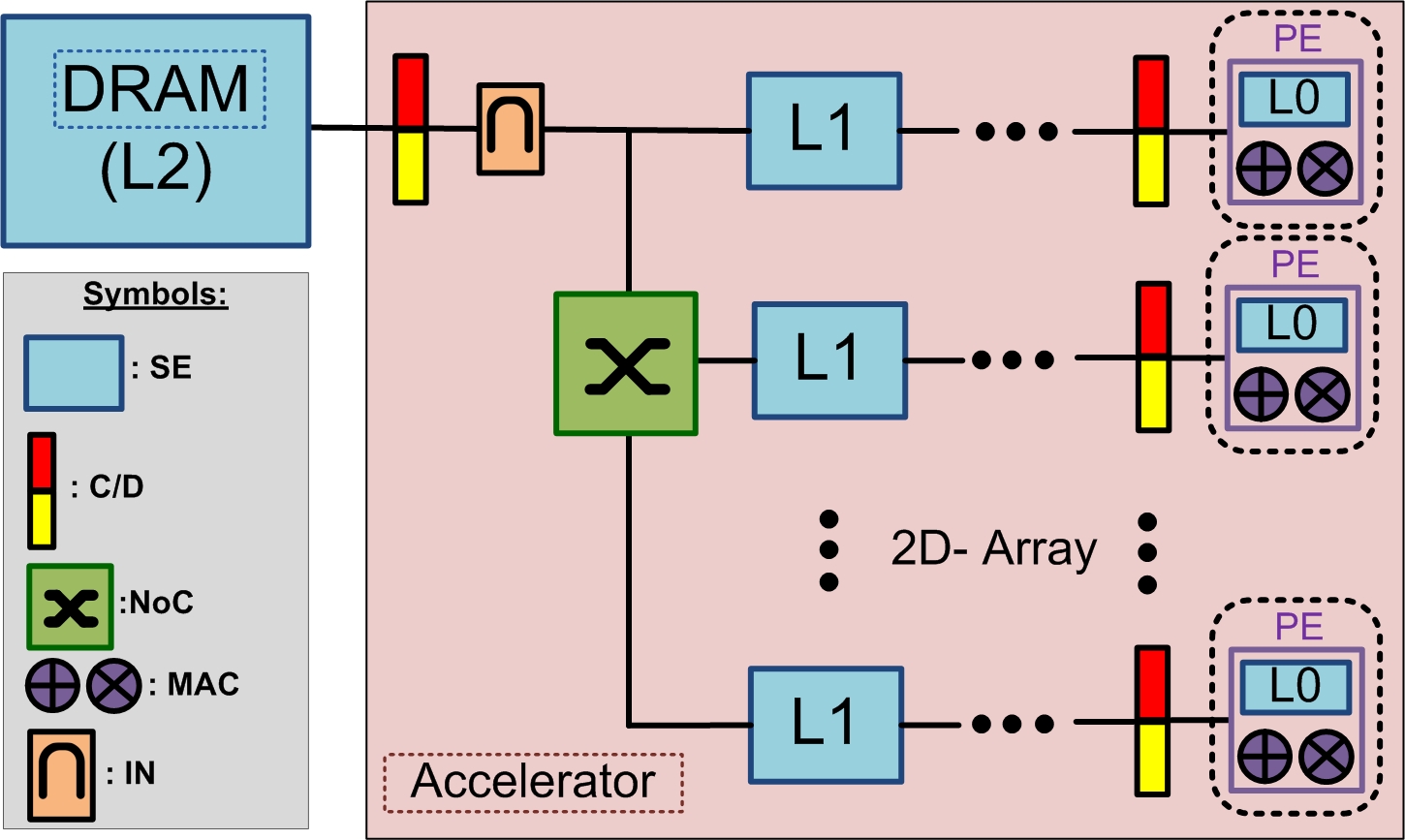}}
\caption{The conceptual design of the row-wise product-based accelerator. 
Abbreviations: Storage Element (SE), CSR compressor and decompressor (C/D), Network on Chip (NoC), Multiply and Accumulate element (MAC), Intersection hardware (IN), Processing element (PE).}
\label{fig.2}
\end{figure}

All row-wise product based accelerators use CSR format, and \textit{CSR compressor/decompressor} units can be placed at each memory level. 
As a general example, Fig. \ref{fig.2} shows CSR compress/decompress units can be located between levels $L_2$ and $L_1$ as well as between $L_1$ and $L_0$.

Another key element is the \textit{intersection logic}, which can be used at different levels of Gustavson-based accelerators. The intersection logic identifies matching non-zero values that must be multiplied from each of the two input matrices. For example, Extensor utilizes intersection hardware between DRAM ($L_2$) and $L_1$ SEs, and in Matraptor, this hardware is located between SpAL and SpBL. The intersection logic is denoted $\cap$ in Fig. \ref{fig.2}.

The processing core of each tensor accelerator is the MAC unit located in each PE. For example, Matraptor has one MAC unit with sorting queue buffers in each PE that operates like a one-dimensional systolic array.
% I don't understand. How many MAC units are there in the Matraptor PE? If the PE is a one-dimensional systolic array, does it not have more than one MAC per PE?
Extensor and G\textsc{amma} use one MAC inside each PE.

All the components of an accelerator are connected through the \textit{Network-on-Chip (NoC)} communication infrastructure. For example, Extensor uses an NoC with unicast, multicast, and broadcast capabilities. Matraptor and G\textsc{amma} employ a customized and simplified crossbar to reduce area overhead.

% --------Maple Processing Element----------
\section{Maple Processing Element}
\begin{figure}[]
\centering
\centerline{\includegraphics[scale=0.45]{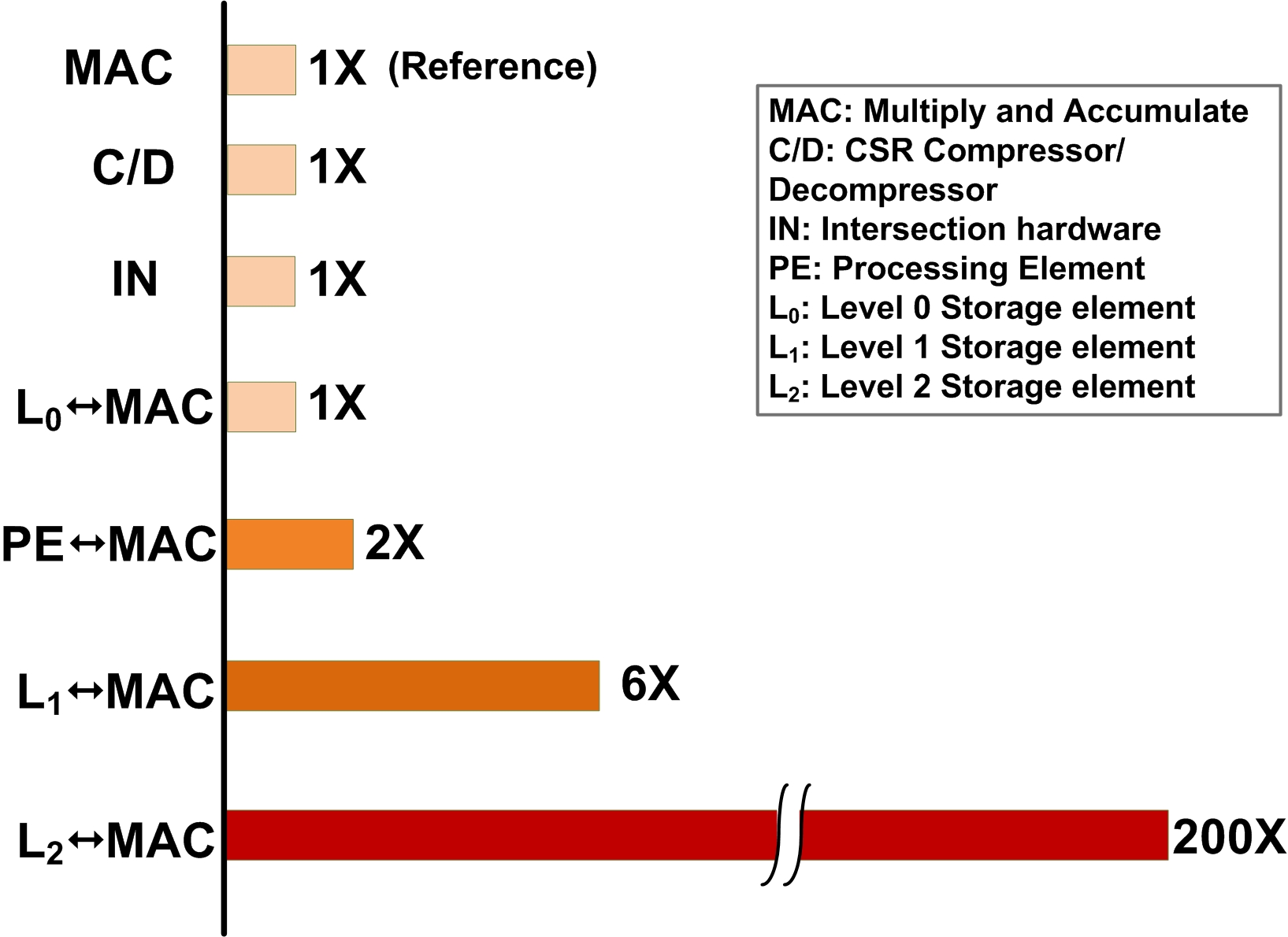}}
\caption{Normalized energy cost of computation and data movement based on 45nm technology extracted from Accelergy. MAC, C/D and IN are computations and $L_0\leftrightarrow MAC$, $PE\leftrightarrow MAC$, $L_1\leftrightarrow MAC$, and $L_2\leftrightarrow MAC$ are data movement operations. $L_i\leftrightarrow MAC$ means data movement between $L_i$ (storage element at $i^{th}$ level) and $MAC$ unit, for instance $L_0\leftrightarrow MAC$.}
\label{fig.3}
\end{figure}

To gain a better understanding of the key factors in sparse tensor accelerators, we measured the energy consumption by several key operations. 
%We extracted the amount of energy consumed by computing operations and data movement in the sparse tensor accelerator to gain a better understanding of the influencing factors in energy efficiency.
%The results are based on 45nm technology obtained from Accelergy \cite{wu2019accelergy}\cite{accelergyweblink}, which is open source and benefits from CACTI\cite{li2011cacti} and Aladdin\cite{shao2014aladdin} plugins.
We use Accelergy \cite{wu2019accelergy}\cite{accelergyweblink} to estimate energy usage, which benefits from CACTI \cite{li2011cacti} and Aladdin \cite{shao2014aladdin} plugins.
It is clear from Fig. \ref{fig.3} that arithmetic consumes less energy than data movement, especially data movement from lower levels of the memory hierarchy. Thus, our primary design goal with Maple is to reduce data movement and improve locality.

% Thus, it is concluded that reducing data movement at the high-level storage and performing processing locally leads in improved energy efficiency.
% As a result, we use this assumption as a foundation for designing Maple processing element in order to enhance energy efficiency.

%  thus increasing the number of PEs will boost the speedup. However, energy efficiency as a primary aim of accelerators should not be overlooked in the name of speedup.

% The Maple processing element made up of several MAC units, to increase parallelism and perform processing locally utilizing the PE's local storage.
% Maple improves more energy efficiency by reducing data movement with higher level storage elements and performing MAC operations locally inside PEs.
% Processing elements are essential components to achieving parallel processing in accelerators. 
To present Maple, we first describe Gustavson's algorithm for sparse matrix multiplication. As shown in Fig. \ref{fig.4}, Gustavson's algorithm (also known as the \textit{row-wise product} approach) consists of two basic operations: \textit{multiply} and \textit{accumulate}.

\textbf{Multiply.} Given that in Gustavson's algorithm, all rows of matrix \textbf{A} are multiplied by rows of matrix \textbf{B}, the resulting partial sums are generally described as:
\begin{equation}
C^k[i,:]=A[i,k]\times B[k,:]    
\end{equation}

Fig. \ref{fig.5}(a) shows an example of partial sums for the first row of matrix \textbf{A} ($A[0,:]$) which are $C^0[0,0]$, $C^0[0,2]$, and $C^2[0,2]$. 

\textbf{Accumulate.} The accumulation operation calculates the matrix \textbf{C} by adding together all partial sums ($C^k[i,j]$): 
\begin{equation}
%C[i,j]=\sum_{0}^{k-1}C^{k}[i,j] old  
C[i,j]=\Sigma_{k} C^{k}[i,j]
\end{equation}

For example, in Fig. \ref{fig.5}(a), $C[0,2]$ is the sum of $C^0[0,2]$ and $C^2[0,2]$.

Eq. (1) to (2) define the row-wise product approach on uncompressed matrices. However, the Gustavson-based technique can also be defined entirely based on the CSR format. 
As a result, the multiply and accumulate operations in CSR representation are defined as follows:

\begin{figure}[]
\centering
\centerline{\includegraphics[scale=0.37]{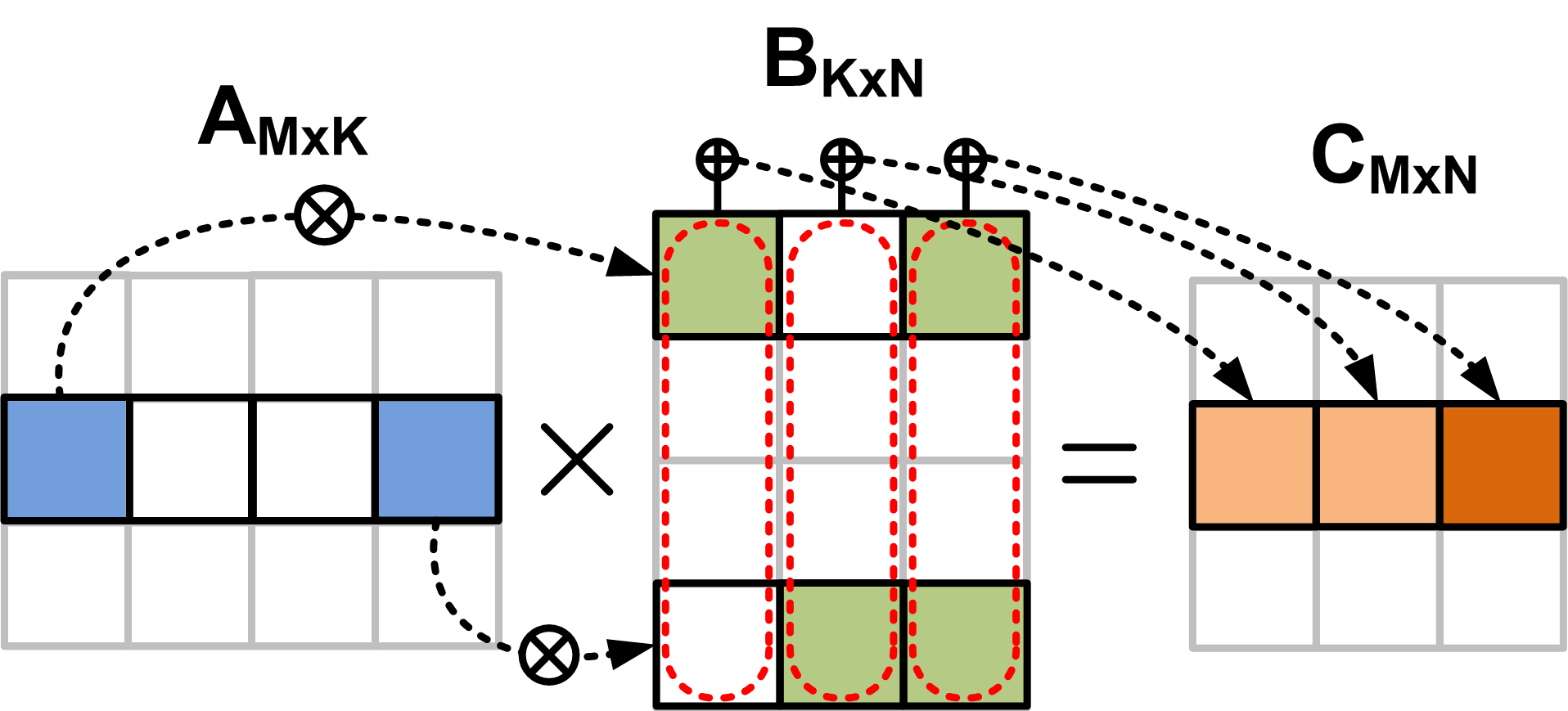}}
\caption{Multiply and accumulate operations in row-wise product-based matrix multiplication.}
\label{fig.4}
\end{figure}

\textbf{CSR-based multiplication.} In the CSR-based multiplication, only non-zero elements of matrices \textbf{A} and \textbf{B} are multiplied, and the metadata  of each matrix specifies the location of non-zero elements during row-wise multiplication.
Indices in the uncompressed matrix serve as the foundation for CSR metadata definition. Therefore, we describe the relation between uncompressed matrix indices and CSR metadata.

For the sake of simplicity, we use the indices $k'$ and $j'$ in CSR-based multiplication, then we define them in detail.

In order to demonstrate the multiplication stage in the CSR format, we extend Eq. (1) as:
\begin{equation}
C^{k'}.value[i][j'] = A.value[i][k']\times B.value[k'][j']
\end{equation}

Where, $A.value[i][k']$ is the non-zero element of $i^{th}$ row and $k'^{th}$ column in the matrix \textbf{A} and $C^{k'}.value[i][j']$ is the $k'^{th}$ partial sum in $i$ and $j'$ locations. Similarly, $B.value[k'][j']$ signifies the non-zero elements of a  matrix \textbf{B} in $k'$ and $j'$ locations.
For example, $C^0.value[0][0]$, $C^0.value[0][2]$, and $C^2.value[0][2]$ are partial sums for the first row of matrix \textbf{A} ($A[0,:]$) in the Fig. \ref{fig.5}(b).

As mentioned earlier, the row index of the matrix \textbf{A} is denoted by $i$.
A significant advantage of the CSR data structure is it allows us to define $k'$ and $j'$ based on $i$ in a row-wise product approach.
As stated in CSR definition in section II.B., $k'$  indices in  matrix \textbf{A} is obtained from $col\_id$ array:
\begin{equation}
k'\gets A.col\_id[i] 
\end{equation}

Since the column index of a non-zero element in matrix \textbf{A} is represented by $k'$, and the column index is indicated by the $col\_id$ array in the CSR data structure. As a result, the column indices of non-zero elements of matrix \textbf{A} are stored in $A.col\_id[i]$ array.
In the same way, $j'$ is obtained from:
\begin{equation}
j'\gets B.col\_id[k']
\end{equation}

Which means the column indices of non-zero elements of matrix \textbf{B} are stored in $B.col\_id[k']$ array. By extending the $k'$ of Eq. (5), $j'$ is obtained from:
\begin{equation}
j'\gets B.col\_id[A.col\_id[i]]
\end{equation}

Eq. (4) and (6) indicate how $k'$ and $j'$ are obtained based on $i$ index. 

\textbf{Accumulate operation based on the CSR format.} The CSR-based accumulate operation is an extension of Eq. (2) and is defined as follows: 
\begin{equation}
% C.value[i][j']=\sum_{0}^{k'-1} C^{k'}.value[i][j'] %old
C.value[i][j']=\Sigma_{k'} C^{k'}.value[i][j']
\end{equation}

Where $C.value[i][j']$ is the final sum produced by the summation of all partial sums.

Maple implements all the mentioned operations at the hardware level. 
Fig. \ref{fig.6} depicts a simple example with four MAC units, where each MAC comprises multiply and accumulate logic. The number of MACs per PE may be determined during the design phase.
% In the next sentence, is a buffer a FIFO? Or something else? I'm having trouble parsing this sentence. For example, it could mean: "According to Eq. (4), the multiply operation requires two buffers: a buffer to store non-zero elements of each rows of \textbf{A} ($A.value[i]$) and another buffer for non-zero elements of the selected rows of \textbf{B} ($B.value[k']$)."

\begin{figure}[]
\centering
\centerline{\includegraphics[scale=0.33]{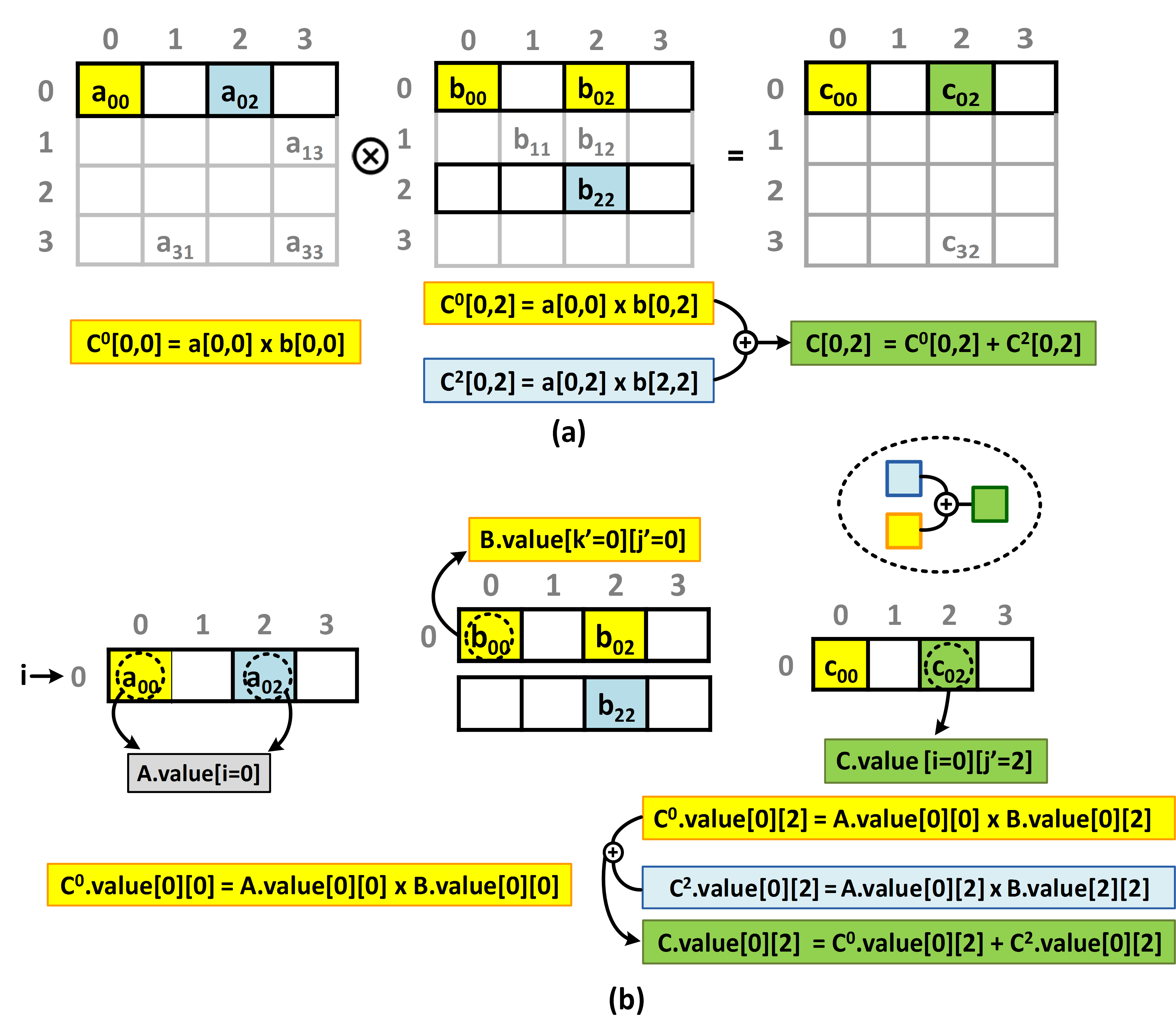}}
\caption{(a) Multiply and accumulate operations on the first row of matrix \textbf{A} with the corresponding two rows of matrix \textbf{B} in the uncompressed matrix presentation. (b) The corresponding CSR components in the compressed matrix presentation. In both figures (a and b), we use color combination (\colorbox{yellow}{\textbf{yellow}} + \colorbox{blue}{\textcolor{white}{\textbf{blue}}} = \colorbox{green}{\textbf{green}}) to illustrate the concept. }
\label{fig.5}
\end{figure}
According to Eq. (3), the multiply operation requires two FIFO buffers to store non-zero elements of each rows of \textbf{A} ($A.value[i]$) and non-zero elements of the selected rows of \textbf{B} ($B.value[k']$). Therefore,  multiply logic consists of two buffers: matrix \underline{\textbf{A}}  \underline{r}ow \underline{b}uffer (ARB) and matrix \underline{\textbf{B}} \underline{r}ows \underline{b}uffer (BRB). 
Fig. \ref{fig.6} shows the placement of non-zero elements of two $4\times4$ matrices in the \textit{\textit{ARB}} and \textit{\textit{BRB}} buffers. 
% This next sentence is a bit unclear. If we multiply a matrix of MxK times a matrix KxN, we get a result matrix MxN. When computing this output matrix of size MxN, we compute MxNxK partial sums. When you write "each row of the matrix, I guess you mean something that is not *every* row of the matrix. Do you mean one row of the matrix, that is a row of size 1xK? 
According to Eq. (1), multiplying each row of the matrix ${A_{M\times K}}$ by matrix ${B_{K\times N}}$ produces $K\times N$ partial sums in row-wise product approach, since each $a_{ik}$ element in matrix \textbf{A} is multiplied by $N$ elements of row $k^{th}$ in matrix \textbf{B}.

Thus, the Maple's accumulate logic utilizes $N$ registers arranged in $1\times N$ dimension called \underline{p}artial \underline{s}um \underline{b}uffer (\textit{PSB}) to store computed partial sums of the multiply logic. 
% We use $PSB[i]$ notation in addressing the \textit{PSB} buffer.
For example, $PSB[0]$ points to the first ($0^{th}$) PSB register.
The outputs of multiply logic are stored in partial sum buffer (PSB) with the following address mapping: 
\begin{equation}
PSB[j'] \gets C^{k}.value[i][j']
\end{equation}

\begin{figure}[]
\centering
\centerline{\includegraphics[scale=0.45]{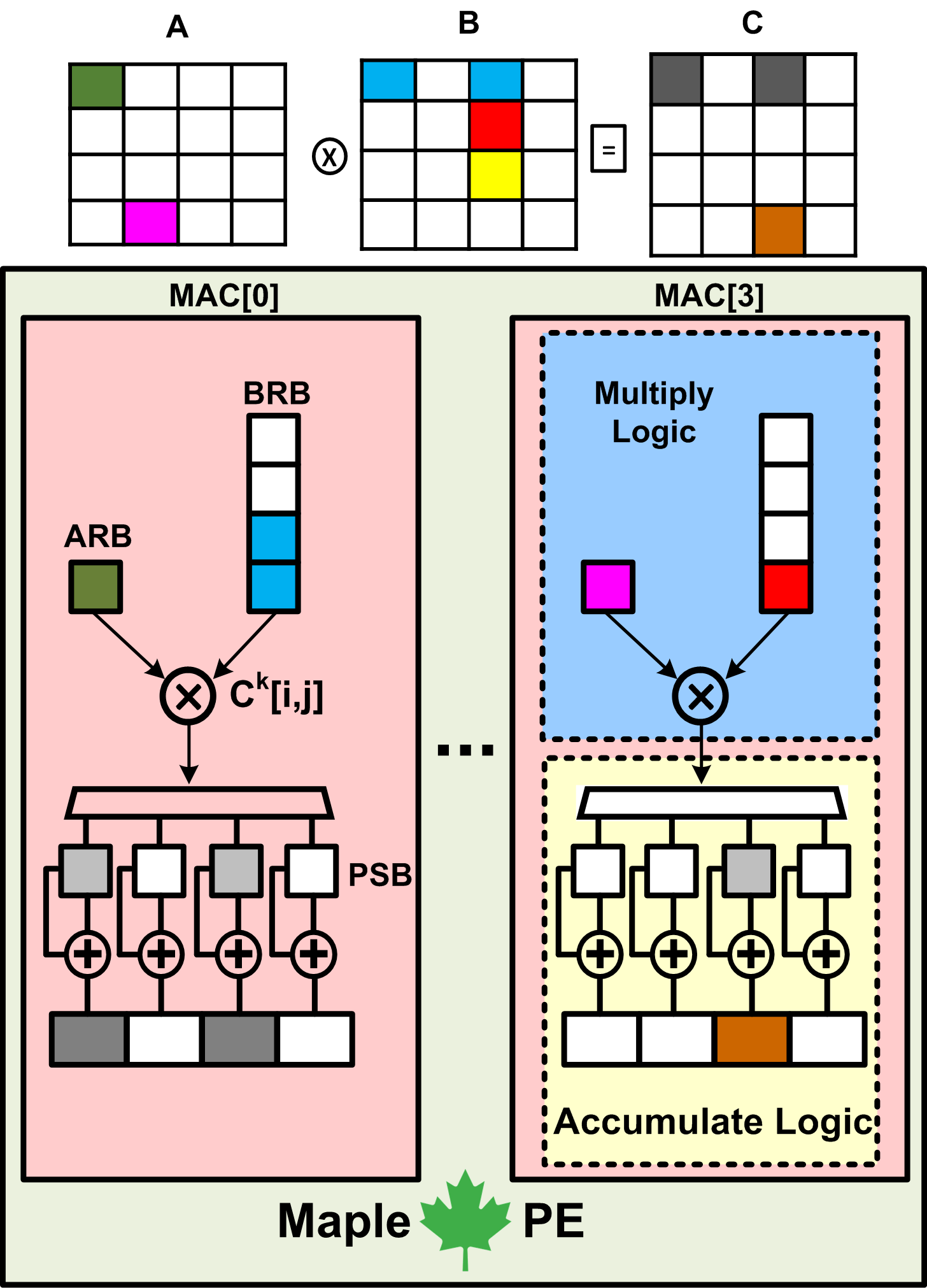}}
\caption{The datapath of a Maple processing element comprising four MAC units for Gustavson-based multiplication of two $4\times4$ sparse matrices \textbf{A} and \textbf{B}.}
\label{fig.6}
\end{figure}

Where, $j'$ denotes the $j'^{th}$ register of \textit{PSB} buffer.
Each $PSB[j']$ is connected to an adder that computes the row of final sums by adding $C^k$ generated partial sums from multiply logic in accordance with Eq. (7).

\begin{figure}[]
\centering
\centerline{\includegraphics[scale=0.45]{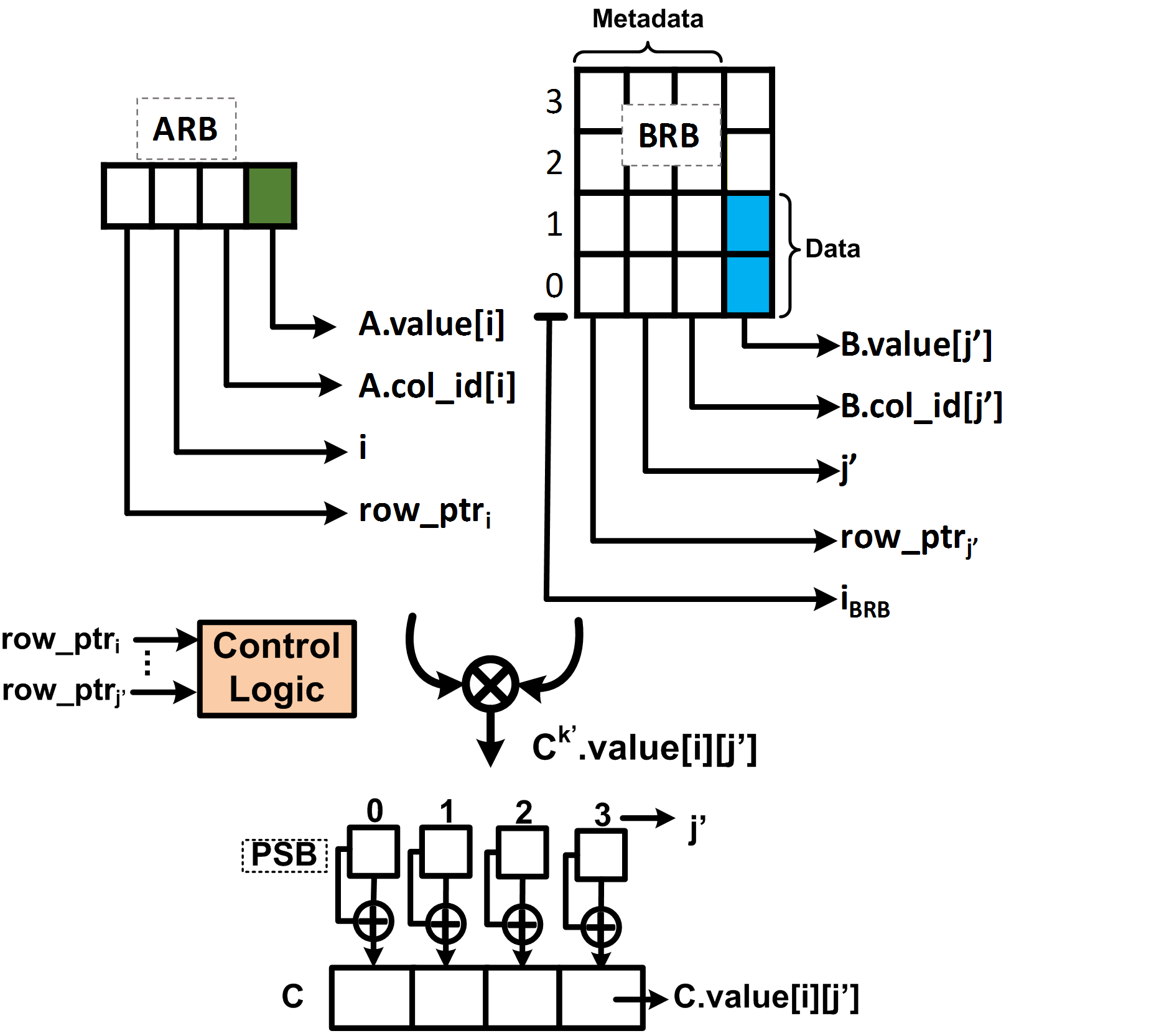}}
\caption{Buffer organization of ARB, BRB and PSB to store nonzero elements and related metadata. The control logic counts the number of multiplications using the $row\_ptr_i$ of the matrix \textbf{A}.} 
\label{fig.7}
\end{figure}

Fig. \ref{fig.7} depicts the details of \textit{ARB}, \textit{BRB}, and \textit{PSB} structures, along with the parts of the buffer that store non-zero elements and associated metadata.

As Fig. \ref{fig.7} shows, $A.col\_id[i]$, $row\_ptr_i$, and, $i$ are metadata for $A.value[i]$ values. Furthermore, we use $j'$ index to simplify the presentation of \textit{BRB} structure. However, $j'$ is defined based on $i$, according to Eq. (6). 

In the CSR data structure, {\fontfamily{lmss}\selectfont row\_ptr} points to the first non-zero element of each row. By subtracting adjacent elements of {\fontfamily{lmss}\selectfont row\_ptr} we can find the number of non-zero values in each row of the matrix. The control logic of the MAC unit uses {\fontfamily{lmss}\selectfont row\_ptr} values to control the number of multiplications of each non-zero elements of matrix \textbf{A}.   

% -----------------Data Level solution---------------------------
\section{Evaluation}
\subsection{Dataset and Simulation}
To evaluate the energy benefit, we employ the Maple processing element in two reference accelerators which operate based on the CSR compression format: Matraptor and Extensor. We do not include G\textsc{amma} in the comparison because it operates on a sparse coordinate format instead of compressed sparse row (CSR). The simulation dataset obtained from SuiteSparse \cite{suitesparse} which covers the variety of matrix sizes and density ranges as shown in Table I. Since Matraptor and Extensor evaluate the performance by multiplying a sparse matrix with itself ($C= A\times A$), we used the same approach to perform a fair comparison.
Moreover, we use the Sparseloop-Accelergy \cite{wu2022sparseloop} toolchain to simulate the baseline and Maple-based configurations. Sparseloop uses Accelergy to evaluate the energy consumption.

\subsection{Main Result}
In this section, we define the baseline and Maple-based configurations for the Matraptor and Extensor.

\subsubsection{Matraptor}
The baseline Matraptor uses two memory levels: SpAL/SpBL ($L_1$), and each PE sorting queues ($L_0$). Additionally, eight processing elements, each with a MAC unit and queue buffers are connected to DRAM via a crossbar.

The Maple-based Matraptor consists of one memory level: Maple's internal buffers (ARB, BRB, and PSB) as $L_0$. Four PEs connect to DRAM via a simplified crossbar similar to the baseline configuration. Each PE consists of two MAC units. Hence, we compare two different configurations with eight MAC units.
\subsubsection{Extensor}
The baseline Extensor comprises two memory levels: LLB and POB buffers ($L_1$), and PEB ($L_0$). Likewise, a NoC interconnection connects 
% DRAM to 
128 processing elements that are arranged in a $16\times 8$ array. Each processing element includes a MAC unit and a PEB buffer.

The Maple-based Extensor has two memory levels including LLB ($L_1$) and Maple's internal buffers ($L_0$). Meanwhile, eight Maple PEs made up sixteen MAC units are connected
% to DRAM 
via NoC. Hence, our comparison is between two configurations with 128 MAC units each.

\subsubsection{Area Benefit}
\begin{figure}[h]
\centering
\centerline{\includegraphics[scale=0.52]{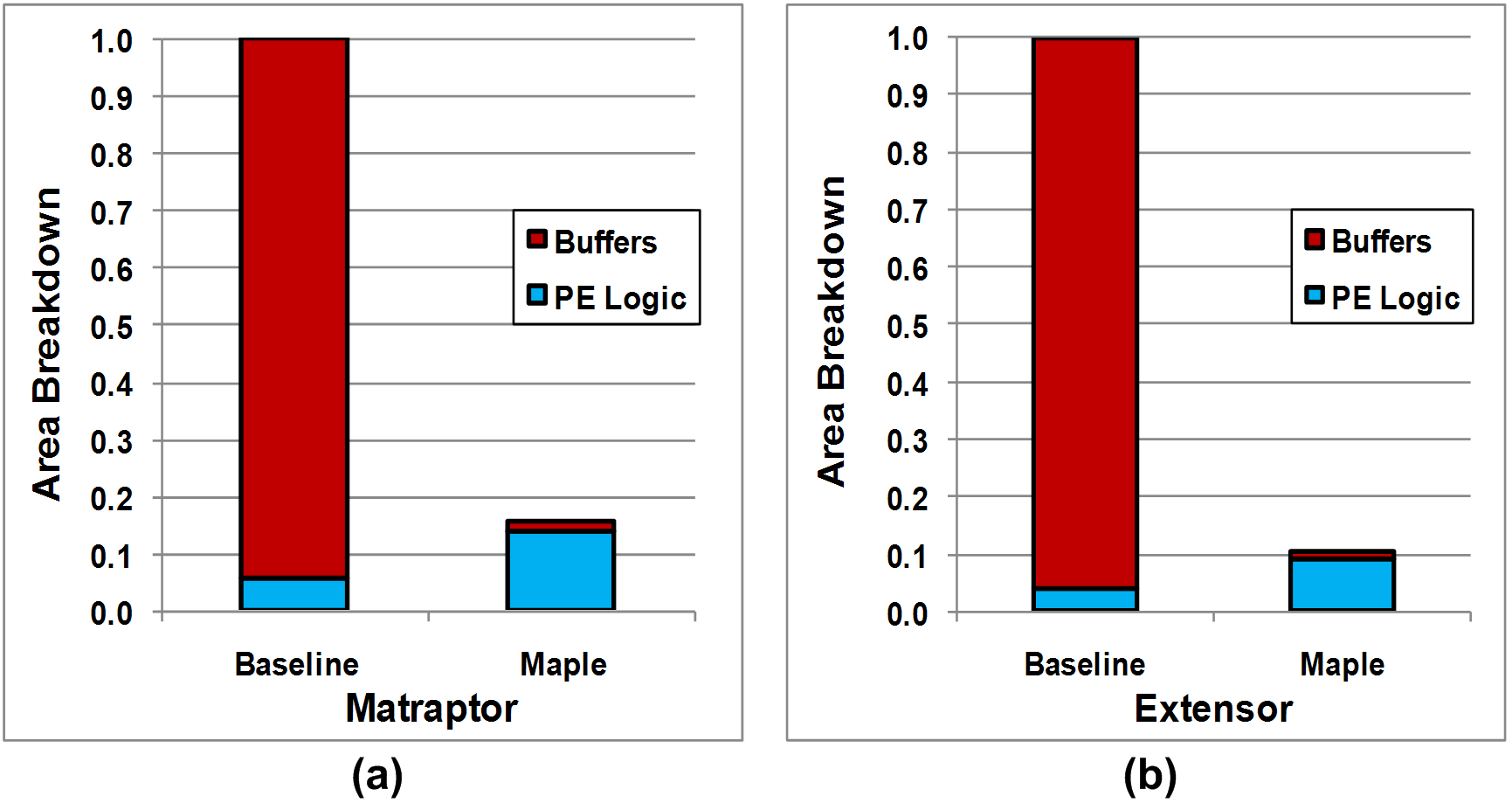}}
\caption{(a) The area consumption of baseline and Maple-based PE in Matraptor, where buffers in Maple are ARB, BRB and PSB. In the baseline Matraptor, sorting queues are PE buffers. (b) The area consumption of baseline and Maple PE in Extensor, where PE's buffers in baseline Extensor are PEB.}
\label{fig.8}
\end{figure}

We evaluate the area consumption of memory elements based on 45nm technology by CACTI 7.0 \cite{muralimanohar2009cacti}.
In addition, we use Aladdin \cite{accelergy-aladdin-plug-in}\cite{Accelergy-Aladdin-Yosys-plug-in} to estimate the area consumption of computational elements. The results also verified by RTL implementation of Maple using Verilog and synthesize by Yosys \cite{wolf2013yosys}\cite{yosys} and the 45nm FreePDK45 library \cite{si2}. 
As shown in Figs. \ref{fig.8}(a) and \ref{fig.8}(b), Maple consumes 84\% and 90\% less area than the baseline PE in Matraptor and Extensor respectively. The reason for the lower area consumption is the use of smaller PE buffers compared to the baseline PE. It is clear that the PEB in Extensor and the PE's sorting queues in Matraptor consume a significant amount of area.  
However, Maple logic consumes the most area because it uses more computational components, such as parallel adders, compared to the baseline PE.

\subsubsection{Energy Benefit}
The energy comparison between the baseline and the Maple-based configurations is shown in Fig. \ref{fig.9}(a). As the results show, the energy consumption in both accelerators is reduced in the Maple-based configuration. The primary reason for energy efficiency is reduced data movement between PE and higher level storage elements. 

Indeed, Maple-based Extensor benefits from computing final results from the partial sum in each PE. Hence, there is no need to utilize POB to store partial sums in a Maple-based configuration. Consequently, the baseline Extensor has a data movement between PE and POB that does not occur in the Maple based Extensor.

In Matraptor, the baseline configuration uses two memory levels, whereas the Maple-based configuration consists of one memory level. Each PE in the baseline Matraptor has one multiply and accumulate logic and functions as a one-dimensional systolic array. The calculation is divided into two phases: generating partial sums from multiply operations and accumulating partial sums through several merge steps. 

Because the baseline PE employs only one multiply and accumulate logic, each PE must use a large sorting queue buffers and conduct the accumulate operation repeatedly in a round-robin fashion.
However, compared to the baseline PE, the Maple base configuration employs several adders in the accumulation logic, which requires less buffer and performs partial sum accumulation in parallel. As a result, the Maple-based configuration achieved higher energy efficiency than the baseline configuration.

\begin{table}[]
\centering
\caption{Simulation Datasets}
\begin{tabular}{llll}
\hline
\multicolumn{1}{c}{\textbf{Matrix}}          & \multicolumn{1}{c}{\textbf{Dim}}       & \multicolumn{1}{c}{\textbf{nnz}}  & \multicolumn{1}{c}{\textbf{Density}} \\ \hline
web-Google (wg)   & 916K$\times$916K & 5.1M & 6.1e-6  \\
mario002 (m2)   & 390K$\times$390K & 2.1M & 1.3e-5  \\
amazon0312 (az)   & 401K$\times$401K & 3.2M & 1.9e-5  \\
m133-b3 (mb)   & 200K$\times$200K & 801K & 2.0e-5  \\
scircuit (sc)   & 171K$\times$171K & 959K & 3.2e-5  \\
p2pGnutella31 (pg)   & 63K$\times$63K & 148K & 3.7e-5  \\
offshore (of)   & 260K$\times$260K & 4.2M & 6.2e-5  \\
cage12 (cg)   & 130K$\times$130K & 2.0M & 1.1e-4  \\
2cubes-sphere (cs)   & 101K$\times$101K & 1.6M & 1.5e-4  \\
filter3D (f3)   & 106K$\times$106K & 2.7M & 2.4e-4  \\
ca-CondMat (cc)   & 23K$\times$23K & 187K & 3.5e-4  \\
wikiVote (wv)   & 8.3K$\times$8.3K & 104K & 1.5e-3  \\
poisson3Da (p3) & 14K$\times$14K   & 353K & 1.8e-3  \\
facebook (fb)   & 4K$\times$4K     & 176K & 1.1e-2  \\ \hline
\end{tabular}
\end{table}

% ---------- Speedup-------------
\subsubsection{Speedup}
\begin{figure}[h]
\centering
\centerline{\includegraphics[scale=0.53]{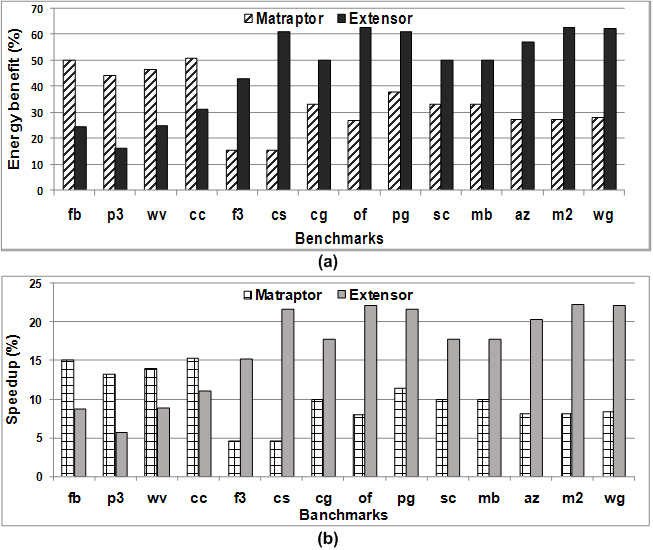}}
\caption{(a) The Energy benefit (\%) of Maple-based Extensor and Matraptor over the baseline structure, (b) The speedup (\%) of Maple-based Extensor and Matraptor over the baseline structure.}
\label{fig.9}
\end{figure}

Fig. \ref{fig.9}(b) shows the Matraptor and Extensor speedup using Maple PE. 
The main reason behind Extensor speedup is the calculation of final sums in each Maple PE without data transfer with POB.

Furthermore, Performing accumulate operation in parallel without multiple repeat is the key reason of speedup in the Maple-based Matraptor configuration. 

\section{Conclusion}
In this paper, we propose the Maple processing element as a basic building block of sparse tensor accelerators. The main goal of using Maple is to minimize data movement between higher level memories that consume a large amount of energy. Maple, operates based on CSR sparse matrices and performs row-wise product approach on nonzero elements by using CSR metadata.
Maple can be used in spatial sparse tensor accelerators consisting of several processing elements. Our evaluation shows that Maple gains large energy benefits in addition to significant speedups. Furthermore, we significantly reduce chip area consumption due to reduced PE-level buffers.

\section{Acknowledgement}
This work was supported, in part, by Science Foundation Ireland under grant No. 13/RC/2094\_P2, co-funded under the European Regional Development Fund through the Southern \& Eastern Regional Operational Programme to Lero \footnote{\href{https://lero.ie/}{https://lero.ie/}} (Science Foundation Ireland Research Centre for Software), and, in part, this project has received funding from the European Union’s Horizon 2020 research and innovation programme under the Marie Skłodowska-Curie grant agreement No 754489.  

\bibliographystyle{IEEEtranN}
\bibliography{main}
\end{document}